# Energy level alignment at $Alq_3$/$La_{0.7}Sr_{0.3}MnO_3$ interface for organic spintronic devices

Y. Q. Zhan*, I. Bergenti, L. Hueso, V. Dediu
Istituto per lo Studio di Materiali Nanostrutturati, Consiglio Nazionale delle Ricerche, ISMN-CNR, via Gobetti 101, 40129 Bologna, Italy

M. P. de Jong
Department of Physics, IFM, Linköping University, S-581 83 Linköping, Sweden

Z.S. Li
ISA, Aarhus University, Denmark

**Abstract**

The electronic structure of the interface between Tris(8-hydroxyquinolino)–aluminum ($Alq_3$) and $La_{0.7}Sr_{0.3}MnO_3$ manganite (LSMO) was investigated by means of photoelectron spectroscopy. As demonstrated recently this interface is characterized by efficient spin injection in organic spintronic devices. We detected a strong interface dipole of about 0.9 eV that shifts down the whole energy diagram of the $Alq_3$ with respect to the vacuum level. This modifies the height of the barriers for the holes injection to 1.7 eV, indicating that hole injection from LSMO into $Alq_3$ is more difficult than it was expected as the energy level matched by vacuum levels. We believe the interface dipole is due to the intrinsic dipole moment characteristic for $Alq_3$ layer. An additional weak interaction is observed between the two materials influencing the N 1s core levels of the organic semiconductor. The presented data are of greatest importance for both qualitative and quantitative description of the organic spin valves.

---

* yq.zhan@bo.ismn.cnr.it

In the past few decades, the field of organic electronics has progressed enormously stimulated by the availability of a virtually infinite number of organic molecules, each with a unique electronic and optical property. While the efforts have been mainly concentrated on the control of charges in various devices, the interest towards spin manipulation in such materials is continuously growing. Due to weak spin-orbit and hyperfine interaction in organic π-conjugated semiconductors, the spin coherence in most organic semiconductors (OS) is expected to be robust and propagate to longer distances than in conventional metals and semiconductors. The first communications on spin injection in OS reported room temperature magnetoresistance in sexithiophene (T6) connected to two manganite ($La_{0.7}Sr_{0.3}MnO_3$) electrodes[1], and real spin-valve effects in Tris(8-hydroxyquinolino) - aluminum ($Alq_3$) confined in a vertical geometry between manganite(bottom) and Co (top) electrodes[2]. Recently spin valve effects have been confirmed for various vertical manganite-$Alq_3$ based devices[3,4]. A spin tunneling across thin $Alq_3$ layer was recently demonstrated to persist up to room temperature[5]. Most of the successful organic devices for spintronic applications were so far based on a combination of $Alq_3$ as OS and manganite thin films as spin polarized injectors. While the bulk and even surface properties of these materials are well understood, the information of their interface, the key region for both charge and spin injection, is completely lacking. A few publications related to interfaces interesting for spintronic applications like C60/Co[6] and pentacene/Co[7,8] are not relevant for our case.

In this letter, we investigate the $Alq_3$/LSMO interface by photoelectron spectroscopy (PES). The samples consist of $Alq_3$ thin films of various thicknesses deposited on LSMO bottom layer. The valence band spectra and secondary electron cut-offs of the $Alq_3$/LSMO interface, core level spectra of both LSMO and $Alq_3$ have been studied in order to obtain the detailed energetics of this interface.

Experiments were carried out using the SX700 beamline at the ASTRID synchrotron source (ISA, Denmark) with photon energies in the 60–600 eV range at room temperature. Electrons were analyzed using a VG CLAM II (30 eV pass energy). Spectra were obtained at normal emission with respect to sample surfaces, and the angle between incident photon beam and direction of the analyzer detection was 45°. The overall resolution was about 0.1 eV. Each spectrum was normalized by the primary photon flux obtained by recording continuously the primary beam intensity on a gold grid. Both valence band spectra and secondary electron cutoffs of $Alq_3$ on LSMO were obtained with a photon energy of 60 eV. The photon energies of 600 eV and 145 eV were used to investigate the N 1s and Al 2p levels of $Alq_3$. Also a bias voltage of $V_b$= -9.4 V was applied to the sample during the secondary binding energy cutoff measurements in order to distinguish between analyzer and sample cutoffs. An evaporation chamber was connected to the analysis chamber, in which the pre-deposited LSMO films were covered *in situ* by a thin $Alq_3$ film. Both analysis and evaporation chambers were baked before the experiments; after baking the background pressures were $8\times10^{-10}$ and $8\times10^{-8}$ mbar respectively.

LSMO films were deposited on matching $NdGaO_3$ substrates (NGO) using the channel spark ablation (pulsed electron deposition)[9]. During the deposition substrates were heated to 800–850 °C, while the oxygen pressure was kept at $10^{-2}$ mbar. This procedure ensures high quality epitaxial LSMO films with high Curie temperature ($T_C$~320–340 K, depending on film thickness) and a resistivity lower than 10 mΩ cm at 300 K. The surface electronic and magnetic properties of our films were characterized in details by various techniques[10--12].

The LSMO substrates (5 mm ×10 mm) were introduced in the evaporation chamber after a rinse with ethanol in an ultrasonic bath. The samples were annealed in UHV and subsequently in a $2\times10^{-5}$ mbar oxygen atmosphere for 30 min at 500 °C. These procedures were found to remove the surface carbon contamination and to restore the surface oxygen stoichiometry[13]. Such oxygen annealing (450–500 °C) is well known and does not change the LEED patterns and the XPS spectra of the manganite film surfaces[14].

The $Alq_3$ films were grown following a step by step sublimation of the organic material from a Knudsen cell at 235 °C. At few chosen thicknesses the deposition was interrupted and the sample transferred to the PES chamber. Subsequently, after PES characterization, the sample is transferred back to the deposition place. The thicknesses were calibrated by deposition time and confirmed later by ex situ Atomic Force Microscopy (AFM) measurements.

In figure 1, the valence band spectra of $Alq_3$ on LSMO are shown as function of its thickness from 0 to 7 nm, where 0 nm of $Alq_3$ corresponds to the investigation of the bare LSMO substrate. Reproducible results were obtained on three investigated samples.

The photoemission spectrum of LSMO (0nm) presents distinct metallic behavior with a broad peak in the 2-8 eV region related to O 2p, and Mn 3d photoelectron emission near the Fermi level [10]. Since LSMO does not have a very sharp Fermi edge, the Fermi edge of Cobalt was used as reference to calibrate the energy scale.

As the $Alq_3$ thickness increases, the emission from the LSMO substrate becomes suppressed and the spectrum continuously changes towards that of $Alq_3$. The 7 nm $Alq_3$ film represents already the typical bulk $Alq_3$ spectrum: seven distinct $Alq_3$ molecular features, in agreement with published reports[15,16], occur at binding energies 2.6 (A), 4.0 (B), 5.0 (C), 7.2 (D), 9.3 (E), 11.4 (F) and 14.5 (G) eV. Structures A and B have been assigned to electron emission from the 2p σ and π orbitals of the 8-quinolinol ligands of $Alq_3$, while the others also preserves the electronic structure of quinolinol [15]. There is almost no contribution from the central Al atom to these orbitals. The $Alq_3$ highest occupied molecular orbital (HOMO) peak corresponds to the feature A which is the nearest peak to Fermi level. The HOMO energy onset is defined by the intersection of the tangent line of the HOMO peak with the baseline of the $Alq_3$ spectrum. In the 7 nm thick film, the HOMO onset of $Alq_3$ is located at $E_{HOMO}$= 1.7 eV below Fermi level; the relative position of the HOMO energy onset from the LSMO Fermi level is thus defined by the comparison of spectra corresponding to 0 nm and 7 nm.

The formation of the interface is described by the valence band evolution at different $Alq_3$ coverage. The spectrum corresponding to 0.06 nm $Alq_3$ is similar to that of the pure LSMO indicating the major contribution of the substrate except a tiny shoulder raised at the position of 2.4 eV. By increasing the thickness to 0.13 nm, the features of $Alq_3$'s valence band becomes evident, especially for the peaks F and G which are not overlapped by the large and broad valence band peak of LSMO. The features A-E also becomes evident by increasing the $Alq_3$ thickness; the whole $Alq_3$ valence band is clearly seen in the spectrum corresponding to an $Alq_3$ thickness of 2.8 nm.

By comparing the peak position of each feature, a significant shift toward lower binding energy with increasing the thickness of $Alq_3$ can be observed. The dashed lines, which connect the peaks of the same feature, are parallel indicating a synchronous shift. By measuring intersections between dashed line of feature G for 0.06 nm and 7 nm $Alq_3$ spectra, an energy shift of about 0.9 eV is found. Since all the peaks in $Alq_3$ valence band are referenced to the Fermi energy of LSMO, the $\Delta = 0.9$ eV shift should be related to a modification of the LSMO work function while depositing $Alq_3$.

The variation of the work function can be directly determined from the secondary electron cutoff energy as function of thickness of $Alq_3$ (fig.2). Let's first analyze the 0 nm curve in fig. 2, corresponding to bare LSMO surface. The work function of the LSMO substrate is calculated as $\phi_{LSMO} = 4.9$ eV in agreement with our previous data[12]. The $Alq_3$ deposition leads to a strong decrease of the work function and, finally, for the thick $Alq_3$ film corresponding to 7 nm thickness, the cutoff is shifted to higher energy by $\Delta_\phi = -0.9$ eV. This value corresponds exactly to the shift found for the valence band features, and unambiguously indicates the presence of a strong interfacial dipole. The origin of such a dipole layer can be attributed to several factors: charge transfer across the interface[17], Pauli repulsion[18,19], strong chemical interaction[20], oriented permanent molecular dipoles[21]. Considering that depositing $Alq_3$ lowers the work function, such that the corresponding interfacial dipole has it's positive pole pointing out of the surface, we ought to exclude a charge transfer mechanism from $Alq_3$ to LSMO because the high ionization potential of $Alq_3$ (5.7 eV) prevents any electron transfer from $Alq_3$ to LSMO. Pauli repulsion can be large for high work function metals featuring a large surface dipole contribution to the work function, but it is expected to be rather small for LSMO. This is because in LSMO the 3d electron density is low compared to that of the 3d transition metals, leading to a correspondingly low electron density leaking out into the vacuum to minimize kinetic energy[22,23]. Hence the surface dipole is expected to be small, and the work function is thus mainly determined by the bulk chemical potential, such that any modification of the surface dipole by Pauli repulsion effects should lead to minor work function changes. Moreover we did not see any evidence of covalent bond formation in the valence band spectra, although there may have been some weak Binding Energy shifts for some elements (shown in figure 4). By far the most likely explanation of the observed dipole is that it stems mainly from the permanent dipole of $Alq_3$, which is rather large: 4 D for the meridional isomer and 7 D for the facial isomer[24].

A simple estimate of the work function change $\Delta\phi$ (in eV) upon adsorbing an areal density n of molecules (in m-2) that each carry a dipole moment $\mu$ (in Cm) can be obtained from the Helmholtz equation, $\Delta\phi = \mu n/(\varepsilon_0 \varepsilon)$, where $\varepsilon$ is the dielectric constant at the interface, determined by the polarizabilities of both Alq3 and LSMO. For a complete, densely packed monolayer, we can estimate the areal density as $2.5\times 10^{18}$ m-2, based on an average lattice constant of $Alq_3$ crystals of about 1 nm (X-ray diffraction data[25]). If we then assume the dielectric constant to be about 4 (a value of 3 is typical for organic semiconductors, while oxides usually have somewhat larger values) an average dipole of 4 D per molecule (the value for the most common meridianal isomer) would give approximately a 1 eV shift, which fits our experimental results very well. Alternatively, depolarization effects might play a role at high coverages. Such effects have been observed in many adsorbate systems, and are traditionally interpreted in the framework of the Topping model[26].

It is worth pointing out that a strong decrease of the work function upon deposition of $Alq_3$ molecules is generally observed, independent of the substrate and it's initial work function. Examples that can be found in the literature include Cu and Au[27], Ag[28], Al and LiF/Al[21,29]. The insensitivity to the specific substrate used indicates that indeed the main contribution to the interfacial dipole must stem from the intrinsic dipole of $Alq_3$. Since the $Alq_3$ molecules are most likely to interact with any substrate through two of it's ligands instead of only one, partial ordering of the molecular dipoles might be expected, although this has not been demonstrated so far.

From figure 1 and 2, we can obtain the electronic structure of the interface between Alq$_3$ and LSMO (figure 3). Based on figure 2, the vacuum level of Alq$_3$ is 0.9 eV lower than the vacuum level of bare LSMO, which is in its turn at 4.9 eV above the Fermi level. The HOMO level of Alq$_3$ is 1.7 eV lower than the Fermi level. Thus, the ionization potential of Alq$_3$ is 4.9-0.9+1.7=5.7 eV, which is similar to the reported value[30].

The energy of the lowest unoccupied molecular orbital (LUMO) edge of the Alq$_3$ layer can be deduced from the energy of the HOMO edge and the HOMO–LUMO splitting. We are interested in a diagram able to describe the charge (spin) injection at this interface. It is quite common to use the optical gap of 2.8 eV [30] to calculate the LUMO energy [28,30,31]. Nevertheless such a definition does not take into account the excitonic binding energy that should somehow increase the real energy of the LUMO level as far as the carrier injection concerns. The most known methods which allow to measure or estimate the single particle band gap are the inversed photoemission spectroscopy (IPES), Scanning Tunneling Spectroscopy (STS) and transport measurements (IV curves), although the later requires the exact knowledge of the transport mechanism. The IPES techniques give quite high values for the single particle gap – up to 4.6-5.2 eV[32]. Such a high value is in strong contradiction with most transport characterizations of the Alq3 based OLEDs [33], and could be caused by the sample modification under the strong flux of electrons. The STS measurements of the empty states provide, on the other hand, a completely non-disturbative method, as it operates at vanishingly low currents (10$^{-12}$ A). A direct STS measured HOMO–LUMO splitting (2.96±0.13 eV) has been reported by S. F. Alvarado[34,35]. This value was confirmed to describe well the charge injection barrier in light emitting diodes [33]. The absolute value of the LUMO level on our diagram can thus be calculated as 2.74±0.13 eV. It provides the 1.26±0.13 eV barrier height for the electron injection (from Fermi level to LUMO level of Alq$_3$) in good agreement with literature data [33] and with our own calculations ( 1 eV )[4].

In according to this diagram, the hole injection barrier is much larger than it would have been expected considering the vacuum level alignment[2,3]. On the other hand, electron injection barrier is smaller. The possibility of electron injection should certainly be considered in those devices involving the LSMO/Alq$_3$ interface. This information is of the great importance for understanding the spin valve behavior and could be the key issue for the high spin injection efficiency characteristic for this interface.

In order to explore the possible interaction at the interface of Alq$_3$/ LSMO, a set of XPS measurements for core levels were performed. Figure 4 shows the evolutions of the N (1s) and Al (2p) core levels upon deposition of Alq$_3$ on LSMO. Following the initial Alq$_3$ deposition, the N (1s) component develops at 399.8 eV. Upon increasing the thickness of Alq$_3$, the N (1s) peak is more and more clear and a shift of 0.9 eV to the lower energy is evidenced, in analogy with the shifts observed in the valence band spectra. Since the peak of the very thin Alq$_3$ can be considered as coming from interface while the thick one represents more information of bulk, this shift is in line with the previously discussed interfacial dipole, shifting the N (1s) to higher binding energy by about 0.9 eV. It is very important to note that the core level spectra of all LSMO elements indicated no observable modification upon varying the Alq$_3$ film thickness.

In Al (2p) spectra, on the other hand, a much smaller shift can be evidenced. The Al (2p) component remains at 74 eV until the thickness of Alq$_3$ increases to 0.69 nm. Between 0.69 nm to 7 nm, a 0.1 eV shift is visible. In principle, a shift of all (core) energies relative to the Fermi energy of LSMO is expected as the interfacial dipole builds up resulting from the adsorption of an increasing number of Alq$_3$ molecules. However, the central position of the Al atom in the molecule places it at a different position within the dipole field compared to the N atoms. In additional complication is that the core-level binding energies are to a large extent determined by final state screening effects, due to the strong localization of the core hole. Especially at interfaces, these screening effects can be large. This makes the analysis of the data in terms of changes in the initial state electron distribution, e.g. the ground state interfacial dipole, less meaningful.

The electronic structure of the Alq$_3$/LSMO interface was investigated by means of Photoelectron Spectroscopy. We detected a strong interface dipole of about 0.9 eV that shifts down the whole energy diagram of the Alq$_3$ with respect to the vacuum level. This modifies the height of the barriers for the holes and electrons injection, 1.7 eV and 1.26±0.13 eV respectively. The intrinsic dipole moment characteristic for Alq$_3$ molecules was suggested to be the origin of the interface dipole, in line with previously reported Alq$_3$/metal interfaces. An additional weak interaction is observed between the two materials influencing the N 1s core levels of the organic semiconductor. We believe these results are of greatest importance for the quantitative description of LSMO/Alq3 based organic spintronic devices.

**Figure capture**

Figure 1: Valence band photoelectron spectra at the Alq$_3$/LSMO interface

Figure 2: Secondary electron cutoffs of Alq3 on LSMO

Figure 3: Schematic energy band diagram of the Alq$_3$/LSMO interface. All the values with circle are measured in our experiments; optical gap (2.8 eV) is got from ref. 30; LUMO level (2.74 ±0.13 eV) is deduced by the HOMO and single particle band gap from ref 33,34.

Figure 4: photoelectron spectra of N 1s and Al 2p at the Alq$_3$/LSMO interface

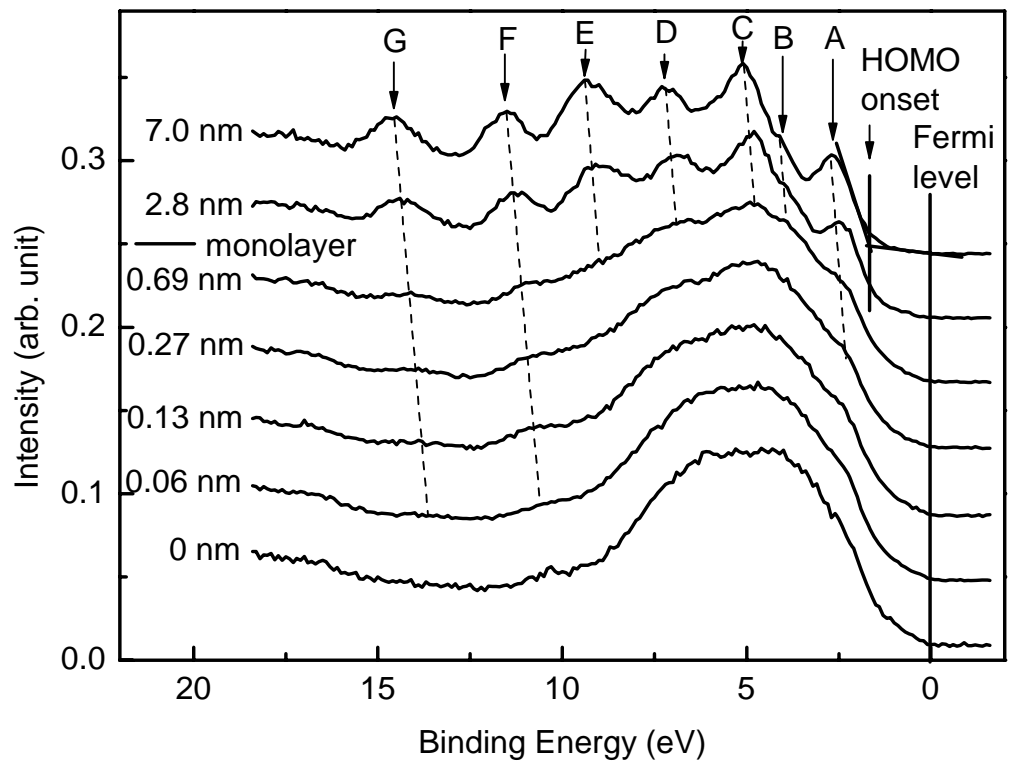

Figure 1

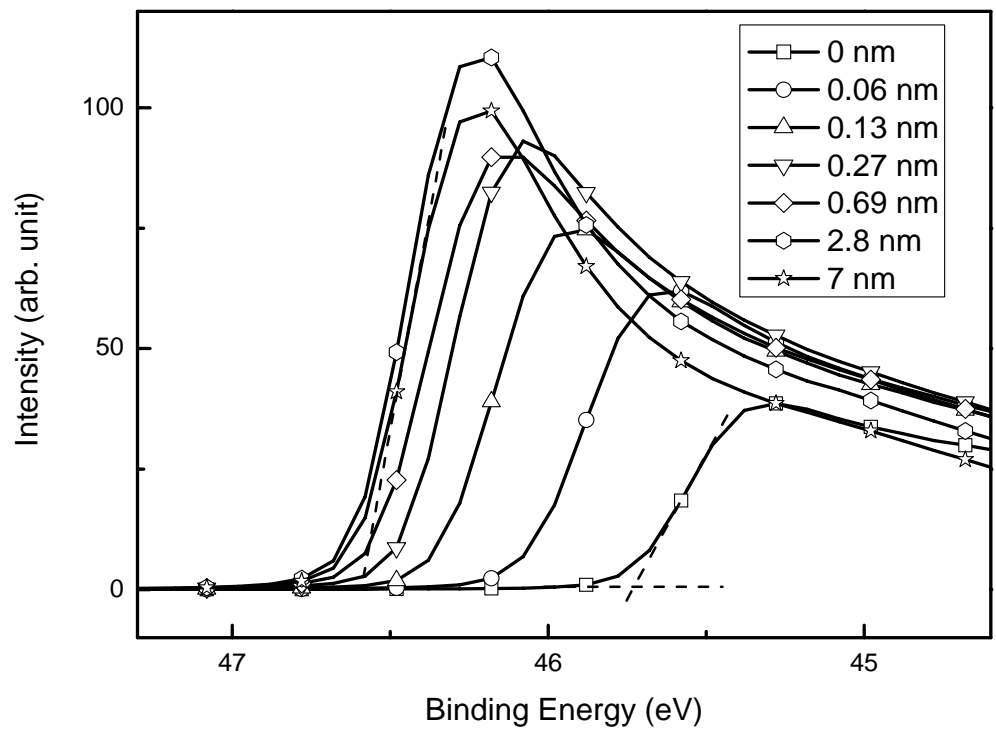

Figure 2

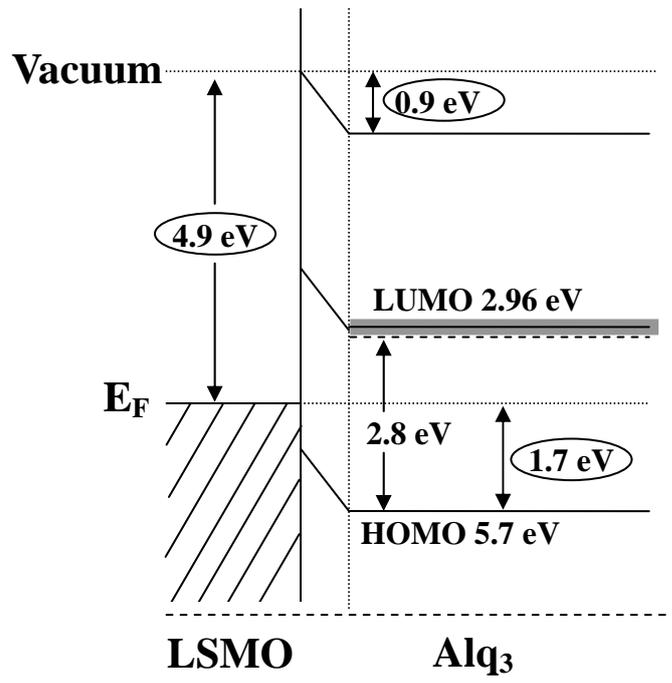

Figure 3

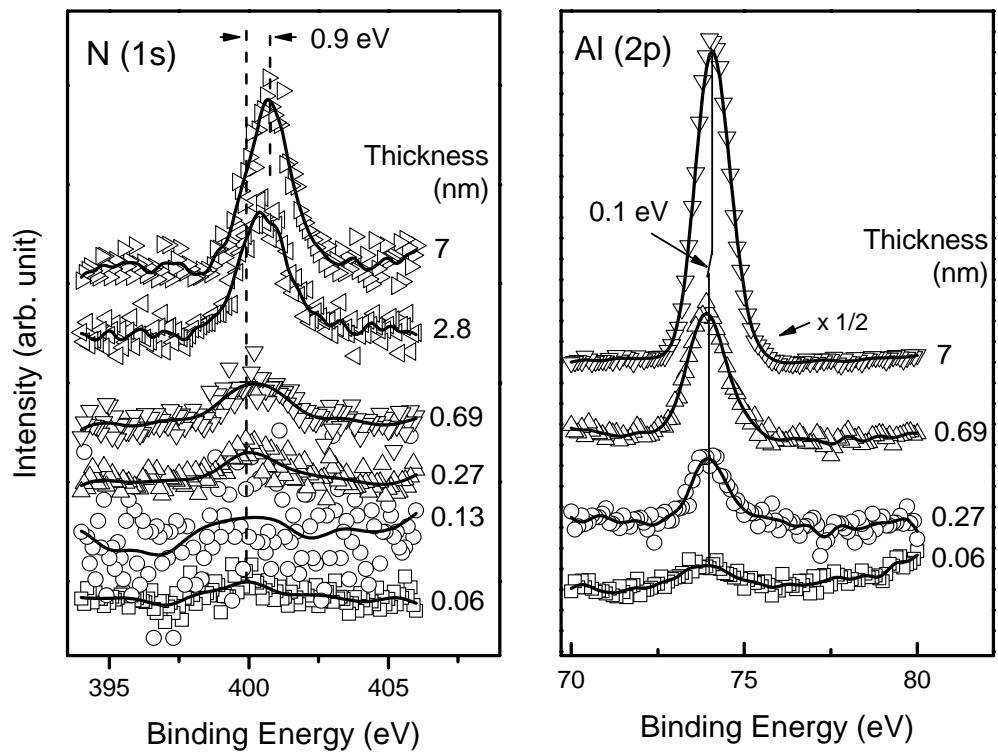

Figure 4